\documentclass[aps,prl,twocolumn,groupedaddress,showpacs]{revtex4-1}
\usepackage[bookmarks]{hyperref}
\usepackage[dvips]{graphicx}
\usepackage{amsmath}
\usepackage{amstext}
\usepackage{graphics}
\usepackage{exscale}
\usepackage{epsfig}
\usepackage{changebar}
\usepackage[T1]{fontenc}
\usepackage{bm}
\usepackage{bbm}

\usepackage{version}%\usepackage{psfrag}

\begin{document}
\title[]{Convergence of energy scales on the approach to a local quantum critical point} 
\author{Y Nishikawa}
\affiliation{Graduate School of Science, Osaka City University, Osaka 558-8585, Japan} 
\author{D J G Crow}
\author{A C Hewson}
\surname{Nishikawa}
\surname{Crow}
\surname{Hewson}
\affiliation{Department of Mathematics, Imperial College, London SW7 2AZ,
  UK.}
\email{a.hewson@imperial.ac.uk}
%\ead{a.hewson@imperial.ac.uk}
%ead{nisikawa@sci.osaka-cu.ac.jp}
%Fermi liquid, NFL and phase transition in model systems
%strongly correlated electrons systems,quantum dots

\date{\today}

\begin{abstract}
We find the emergence of  strong correlations and universality on the approach
to the quantum critical points of a two impurity Anderson model. The  two impurities are
coupled  by an
inter-impurity exchange interaction $J$ and direct interaction $U_{12}$ and are  hybridized
 with separate conduction channels.
The low energy behavior is described in terms of
renormalized parameters, which can be deduced from  numerical
renormalization group (NRG) calculations. We show that on the approach to the
 transitions to   a local singlet and  a local charged ordered state, 
 the quasiparticle weight factor $z\to 0$, and  the
renormalized parameters can be expressed in terms of a single energy scale
$T^*$. The values of the renormalized
interaction parameters in terms of $T^*$ can be predicted from the condition
of continuity of the spin and charge susceptibilities, and correspond to
strong correlation as they are greater than or equal to the effective band width.
These predictions are confirmed by the NRG calculations, including the case when
the onsite interaction $U=0$.
\end{abstract}
\maketitle

There is increasing interest, both experimentally and theoretically,
in strongly correlated electron systems which have anomalous behavior in the region of a $T=0$ or quantum
phase transition (for a recent review see \cite{SS10}). The generalizations of the Wilson renormalization group
approach to zero temperature transitions, initiated by the early  work
of Hertz  and followed up by others \cite{Her76}, 
 have not provided a comprehensive framework
to explain many examples of quantum critical behavior. Particularly challenging
is the range of anomalies observed at quantum critical points (QCP) in heavy
fermion materials, which have been induced by lowering the transition temperature of a
magnetically ordered state to zero by pressure, alloying or in some cases by
an applied magnetic field \cite{SS10,LRVW07,CPSR01}.
\par
One possible mechanism that has been put forward for some heavy fermion systems is that, at the critical point, 
 there is a breakdown of the Kondo
screening such that the associated Kondo resonance disappears \cite{SS10,CPSR01}. This would
imply that the f-like quasiparticles  no longer
contribute the  Fermi surface, so that the Fermi surface would shrink
from a large to a small one, containing only the itinerant non-f electrons.
It is difficult to test this conjecture for a lattice model of this situation
using the theoretical techniques currently available. There is, however, a
 two impurity Kondo
model which has been shown to have a quantum critical point
 \cite{JV87,*JV89,JVW88,SVK93,Gan95,ALJ95},  which might
throw some further light on this particular mechanism. In this Letter, we study a
 related two impurity
 model using a
combination of renormalized perturbation theory (RPT) and numerical
renormalization group (NRG) calculations. Using  these  techniques
we predict that a single energy scale $T^*$ emerges as the critical point is
approached, such that at the critical point $T^*=0$, and the quasiparticle
Kondo resonance disappears. These results lead to a new perspective
on the two impurity model, and evidence supporting the conjectures
that a quantum critical point can be associated with the collapse of the
Kondo quasiparticle resonance.\par
    The Hamiltonian of the model we will study has  the form, 
${\cal H}=\sum_{\alpha=1,2}{\cal H}_\alpha+{\cal H}_{12}$
 where ${\cal H}_\alpha$ corresponds to an  Anderson impurity model
in channel $\alpha$ given by
\begin{eqnarray}
&&{\cal H}_\alpha=\sum_{\sigma}\epsilon_{d,\alpha}d^{\dagger}_{\alpha,\sigma}d^{}_{\alpha,\sigma}+\sum_{k,\sigma}\epsilon_{k,\alpha}
c^{\dagger}_{k, \alpha,\sigma} c^{}_{k, \alpha,\sigma}\label{model1a} \\
&&+\sum_{k,\sigma} (V_{k,\alpha} d^{\dagger}_{\alpha,\sigma} c^{}_{k,\alpha,\sigma}
+ {\rm h.c.})+ U_\alpha n_{d,\alpha,\uparrow}n_{d,\alpha,\downarrow} \nonumber
\end{eqnarray}
where $d^{\dagger}_{ \alpha,\sigma}$, $d^{}_{ \alpha,\sigma}$, are creation and
annihilation operators for an electron at the impurity site in channel
$\alpha$, where $\alpha=1,2$,  and spin
component
$\sigma=\uparrow,\downarrow$.  
The creation and annihilation operators $c^{\dagger}_{k,\alpha,\sigma}$, $c^{}_{k,\alpha,\sigma}$ are
for  partial wave conduction electrons with energy
$\epsilon_{k,\alpha}$ in channel $\alpha$. \par
The second part of the Hamiltonian ${\cal H}_{12}$  decribes the interaction
between the impurities in the two channels, which we take in the form of a
direct Coulomb term $U_{12}$ and  an  Heisenberg exchange term,
\begin{equation}
{\cal H}_{12}= U_{12}\sum_\sigma n_{d,1,\sigma} 
\sum_{\sigma'} n_{d,2,\sigma'}+
2J_{} {\bf S}_{d,1}\cdot{\bf S}_{d,2},  
\label{model2c}
\end{equation}
where $J>0$ for an antiferromagnetic coupling.\par 
For the model with $U_{12}=0$ there  is a competition between two modes of
screening of the impurity spins;  by  Kondo screening in the channel directly hybridized to each
impurity or  by the  direct antiferromagnetic coupling between the
impurities. The Kondo screening by the individual channels predominates for
small $J$, and the local screening for large $J$. 
NRG studies   \cite{JV87,*JV89} of the Kondo version of the symmetric  model
have shown that there is a critical point between these competing terms
at a value $J=J_c$, where $J_c$ is proportional to the Kondo temperature
$T_{\rm K}$ of an
isolated impurity ($J=0$). It was also shown there is a  divergence of the impurity
specific heat coefficient at the transition point and  an anomalous ${\rm log}(2)/2$ entropy.
 Conformal field theory \cite{AL92,ALJ95} and bosonization  studies
\cite{Gan95} have clarified the conditions for such a transition to occur and
shown that
 the staggered
susceptibility diverges at the transition point.\par
 The model with $J=0$, $U_{12}\ne 0$, has been used to describe two capacitively
coupled quantum dots, and NRG studies have revealed a second type of transition at a critical
value of $U_{12}=U^c_{12}$,  such that for $U_{12}>U^c_{12}$ there is a breaking
of local charge order \cite{GLK05,*GLK06}. Further recent NRG studies of both types of
 transitions have been reported using related versions of this model to
 describe the electron transport in 
double quantum dots \cite{CH07,LWG09,FHLS03,MSA10,JGL11}.
 \par
In the renormalized perturbation theory \cite{Hew93,*Hew01}, the impurity retarded Green's
function $G_{d,\alpha,\sigma}(\omega)$  is re-expressed as $G_{d,\alpha,\sigma}(\omega)=z_\alpha\tilde G_{d,\alpha,\sigma}(\omega)$,
where $\tilde G_{d,\alpha,\sigma}(\omega)$ is the quasiparticle Green's function
given by 
\begin{equation}
\tilde G_{d,\alpha,\sigma}(\omega)=\frac{1} {\omega-\tilde\epsilon_{d} +i\tilde\Delta-\tilde\Sigma_{\sigma}(\omega)}\label{qpgf}
\end{equation}
and the renormalized parameters, $\tilde\epsilon_{d}$ and $\tilde\Delta$
are given by
\begin{equation}\tilde\epsilon_{d} =z(\epsilon_{d}+\Sigma_{\sigma}(0)),\quad \tilde\Delta
=z\Delta,\label{rself}
\end{equation}
where $z=1/(1-\partial\Sigma_{\sigma}(\omega,0)/\partial \omega)$
evaluated at $\omega=0$ and
 $\Delta=\pi\sum_k|V_{k}|^2\delta(\epsilon_k)$. We have taken the two channels to be
 equivalent and have dropped the channel index.\par
 In working with the fully renormalized
quasiparticles, it is appropriate to use
the renormalized  or effective interactions between the quasiparticles which
we identify with the   renormalized 
 local four vertices
$z^2\Gamma^{\alpha,\beta}_{\sigma,\sigma',\sigma'',\sigma'''}(\omega_1,\omega_2,\omega_3,\omega_4)$ 
in the zero frequency limit, eg.
$\tilde
U_\alpha=z^2\Gamma^{\alpha,\alpha}_{\uparrow,\downarrow,\downarrow,\uparrow}(0,0,0,0)$ \cite{Hew93,*Hew01}.

%% J=z_1 z_2\Gamma_{{\bf \sigma_1},{\bf \sigma}_2}(0,0,0,0).
The effective Hamiltonian which describes the low energy excitations
corresponds to the original model given in equations (\ref{model1a}) and (\ref{model2c}) with the parameters replaced by
the renormalized values, and the 
 interaction terms have to be normal ordered. For the complete renormalized perturbation expansion to include higher energy scales, counter terms have also to be included to cancel off any further renormalizations \cite{Hew93,*Hew01}.\par 
 
The quasiparticle interaction terms
 do not contribute to the
linear  specific heat coefficient $\gamma$ of the impurities, which is given by
\begin{equation}
 \gamma=4\pi^2\tilde \rho^{(0)}(0)/3,\quad\tilde \rho^{(0)}(\omega)=\frac{\tilde\Delta/\pi}{ (\omega-\tilde\epsilon_{d})^2 +\tilde\Delta^2},
\label{gam}
\end{equation} 
where $\tilde \rho^{(0)}(\omega)$ is the free quasiparticle density of states.
 Exact results for the spin
susceptibility and charge susceptibilities \cite{NCH10s,*NCH10a}, $\chi_s$ and $\chi_c$,
 are given by 
\begin{eqnarray}
\chi_s=4\mu_{\rm B}^2\tilde\eta_s\tilde \rho^{(0)}(0),\quad\tilde\eta_s=1+(\tilde U-\tilde J)\tilde\rho^{(0)}(0),\\
\chi_c=4\tilde\eta_c\tilde \rho^{(0)}(0),\quad\tilde\eta_c=1-(\tilde U +2\tilde U_{12})\tilde\rho^{(0)}(0),
\label{chi}
\end{eqnarray}
and the phase shift $\delta={\rm tan}^{-1}(
\tilde\Delta/{\tilde\epsilon_d})$ per spin per channel.\par

We consider first of all the symmetric model with $U_{12}=0$. When $J=0$ the model corresponds
to two independent Anderson models and the low energy behavior corresponds to
a local Fermi liquid in terms of the two renormalized parameters
$\tilde\Delta$ and $\tilde U$. 

When we switch on and increase an antiferromagnetic coupling $J$  
 this quasiparticle Fermi liquid picture breaks down at a particular coupling
 $J=J_c$, when the quasiparticle weight $z\to 0$ implying  $\tilde\Delta\to 0$. This
in turn implies  that the specific heat
coefficient $\gamma$ given by  Eq. (\ref{gam}) diverges at this point. 
However, as we approach this point we would not expect  the local uniform
spin and charge susceptibilities to diverge as these susceptibilities are
suppressed by an antiferromagnetic coupling and  the critical fluctuations occur
in the staggered spin channel. From
 Eq. (\ref{chi})
this can only be avoided if in turn the coefficients $\tilde\eta_s\to 0$
and $\tilde\eta_c\to 0$  at the same
point, so the product with the singular part remains finite.
Assuming $\tilde U_{12}=0$ these conditions imply

\begin{equation}
\tilde J\to 2\pi\tilde\Delta,
\quad  \tilde U\to \pi\tilde\Delta,\quad{\rm as}\quad J\to J_c. 
\label{prediction1}
\end{equation}
We can define an energy scale $T_<^*$ via $\pi\tilde\Delta(U,J)=4T_<^*$,
such that $T^*_<$ evolves continuously from the value $T_{\rm K}$ for $J=0$.
This implies that on approach to  the critical point the renormalized
parameters can all be expressed in terms of this single energy scale,
$\tilde J/2=\tilde U=\pi\tilde\Delta=4T_<^*$.
%% In the next section we show that from the results of 

We can apply precisely the same arguments to the model with finite $U_{12}$
and $J=0$, which has a local charge ordered transition as $U_{12}(>0)$
is increased to a critical value $U_{12}^c$. Assuming $\tilde J=0$, we find
\begin{equation}
\tilde U_{12}\to \pi\tilde\Delta,
\quad \tilde U\to -\pi\tilde\Delta,\quad{\rm as}\quad U_{12}\to U^c_{12}. 
\label{prediction2}
\end{equation}
 \begin{figure}[!htbp]
   \begin{center}
     \includegraphics[width=0.33\textwidth]{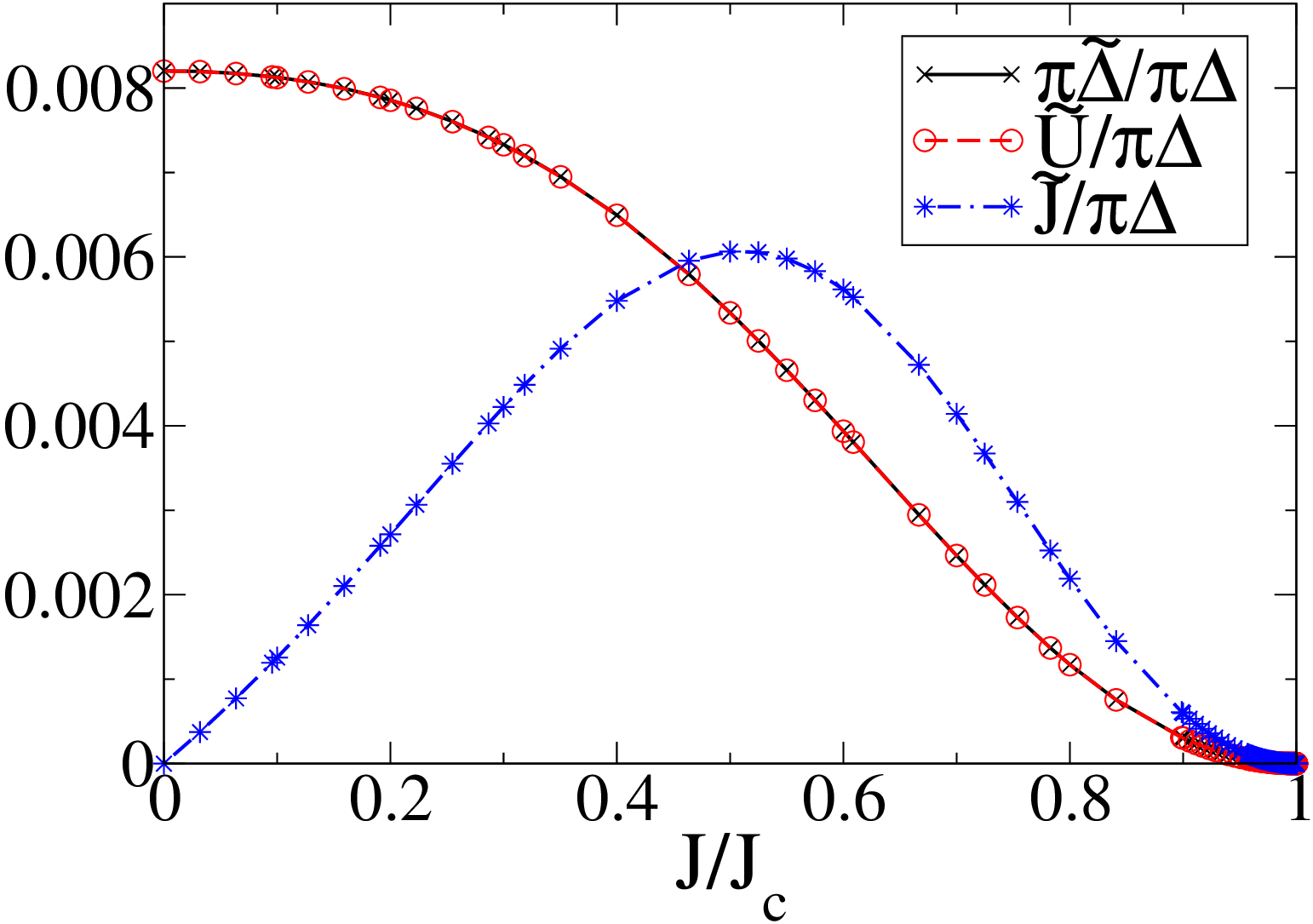}
     \caption{(Color online) A plot of  $\tilde \Delta/\Delta$,  $\tilde U
       /\pi\Delta$ (coincident) and $\tilde J /\pi\Delta$ as a function of $J/J_c$
   for  $U/\pi\Delta=5$, $\pi\Delta=0.01$.
} 
     \label{rp_U5}
   \end{center}
 \end{figure}
Using the RPT approach \cite{NCH10s,*NCH10a} we can calculate the exact asymptotic behavior of the
impurity retarded self-energy $\Sigma(\omega,T)$ for $\omega,T<<T_<^*$ from the second
order calculation of $\tilde\Sigma(\omega,T)$ as
${\rm Im}\,\Sigma(\omega,T)={\rm Im}\,\tilde\Sigma(\omega,T)/z$. The result is
\begin{equation}
\Sigma(\omega,T)=\frac{-i\pi^2I\Delta}{64}\left[\left(\frac{\omega}{
      T_<^*}\right)^2+\left(\frac{\pi T}{ T_<^*}\right)^2\right],
\end{equation}
where $I=(2\tilde U^2+3\tilde J^2+4\tilde U_{12}^2)/(\pi\tilde\Delta)^2$,
so $I\to 14$ as $J\to J_c$ and $I\to 6$ as $U\to U_{12}^c$.\par

\begin{figure}[!htbp]
   \begin{center}
     \includegraphics[width=0.218\textwidth]{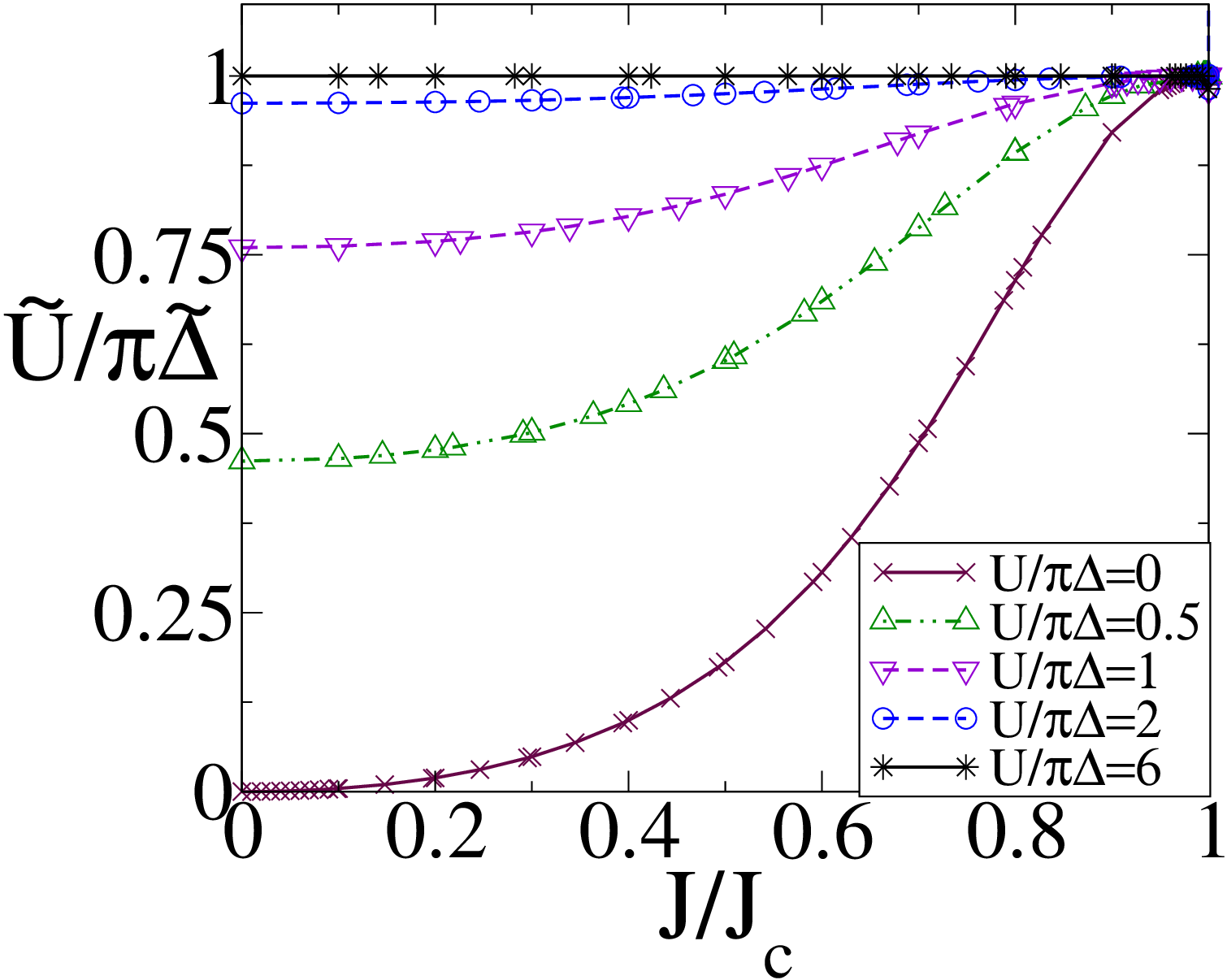}
  \includegraphics[width=0.22\textwidth]{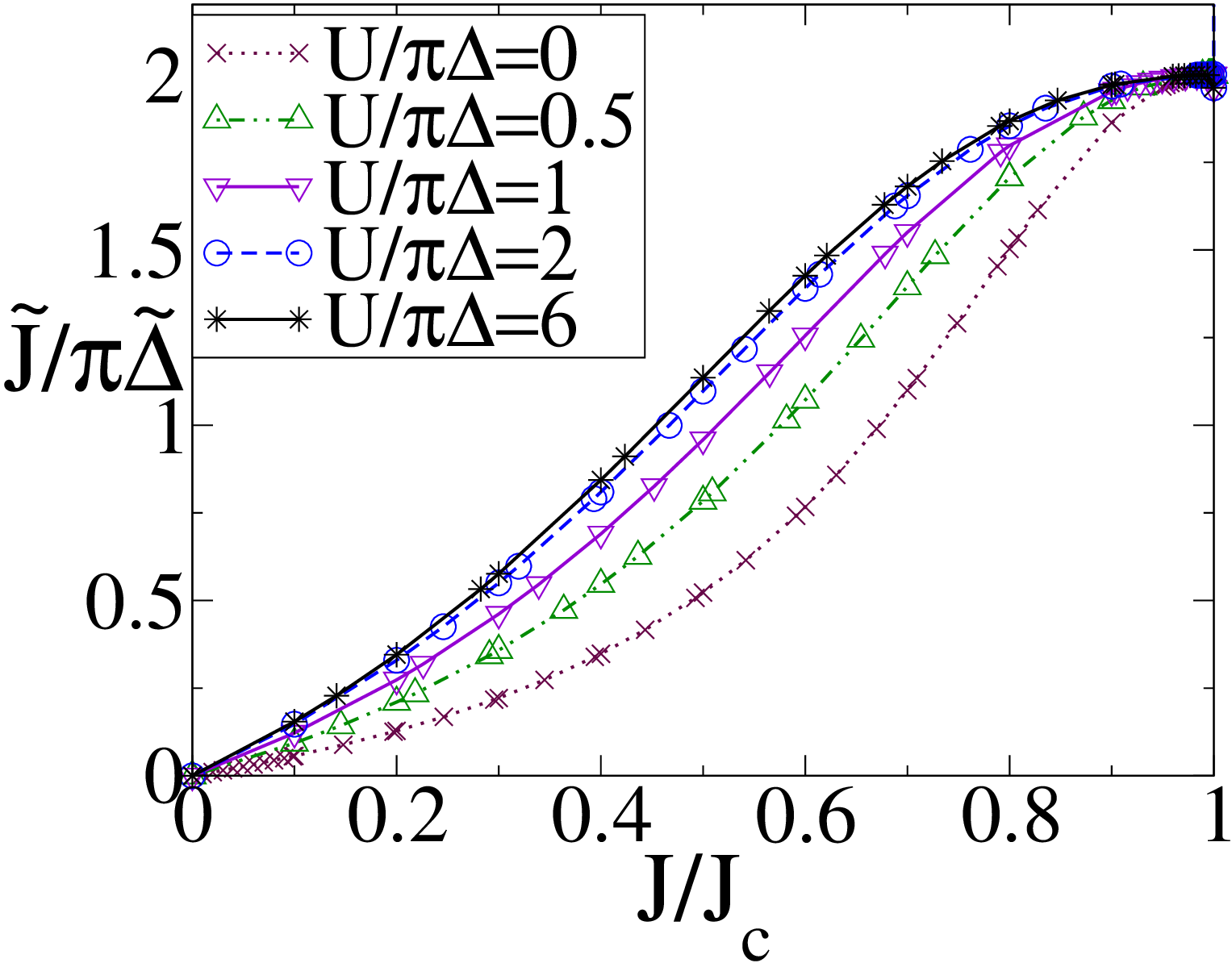}
     \caption{(Color online) A plot of  $\tilde U/\pi\tilde \Delta$ (left)
and  $\tilde J/\pi\tilde \Delta$ (right), as a
       function of $J/J_c$ for  $ U/\pi\Delta=0,0.5,1,2,6$.
} 
     \label{tilu}
   \end{center}
 \end{figure}

We have shown in  earlier work how the renormalized parameters
 can be deduced from an analysis of the low energy fixed point
  of an NRG calculation \cite{HOM04,NCH10s,*NCH10a}, and we apply the same
 procedure for this model to calculate  the low energy behavior and test the predictions given in Eqns.
 (\ref{prediction1}) and (\ref{prediction2}).
In Fig. \ref{rp_U5} we give results for the ratio of $z=\tilde\Delta/\Delta
$, $\tilde U/\pi\Delta$ and $\tilde J/\pi\Delta$ for  $U/\pi\Delta=5$ as a
 function of 
$J/J_c$, where $J_c=1.378T_{\rm K}$. It can be seen that all three renormalized  parameters tend to zero
at the transition point. The fact that $z\to 0$ implies that the Kondo
resonance at the Fermi level disappears at the transition. The ratio of
 parameters, $\tilde U/\pi\tilde\Delta$ and $\tilde J/\pi\tilde\Delta$ remain
 finite.  For $U/\pi\Delta=5$ we are 
in the strong coupling limit of the Anderson model ($J=0$) where $\tilde
 U/\pi\tilde\Delta=1$, and it remains at this value over the whole
$J$ interval. This is not the case in general, however, as can be seen in Fig. \ref{tilu}
 (left) where values of $\tilde U/\pi\tilde \Delta$ are given as a function of $J/J_c$
for $ U/\pi\Delta=0,0.5,1,2, 6$. It can be seen that as $J\to J_c$,
 $\tilde U/\pi\tilde\Delta\to 1$ in all cases {\em including the case} $U=0$, so that
 strong correlation result emerges even in the weak coupling case on the approach to the critical point.
In the right panel of Fig. \ref{tilu}, the corresponding values of 
 $\tilde J/\pi\tilde \Delta$ are shown and all converge to the limiting
value  $\tilde J/\pi\tilde \Delta=2$, which confirms the predictions given in
Eqn. (\ref{prediction1}). The results also apply for the model with $U_{12}\ne
 0$ as it is found that $\tilde U_{12}/\pi\tilde\Delta\to 0$ as $J\to J_c$. 
\par

 \begin{figure}[!htbp]
   \begin{center}
     \includegraphics[width=0.27\textwidth]{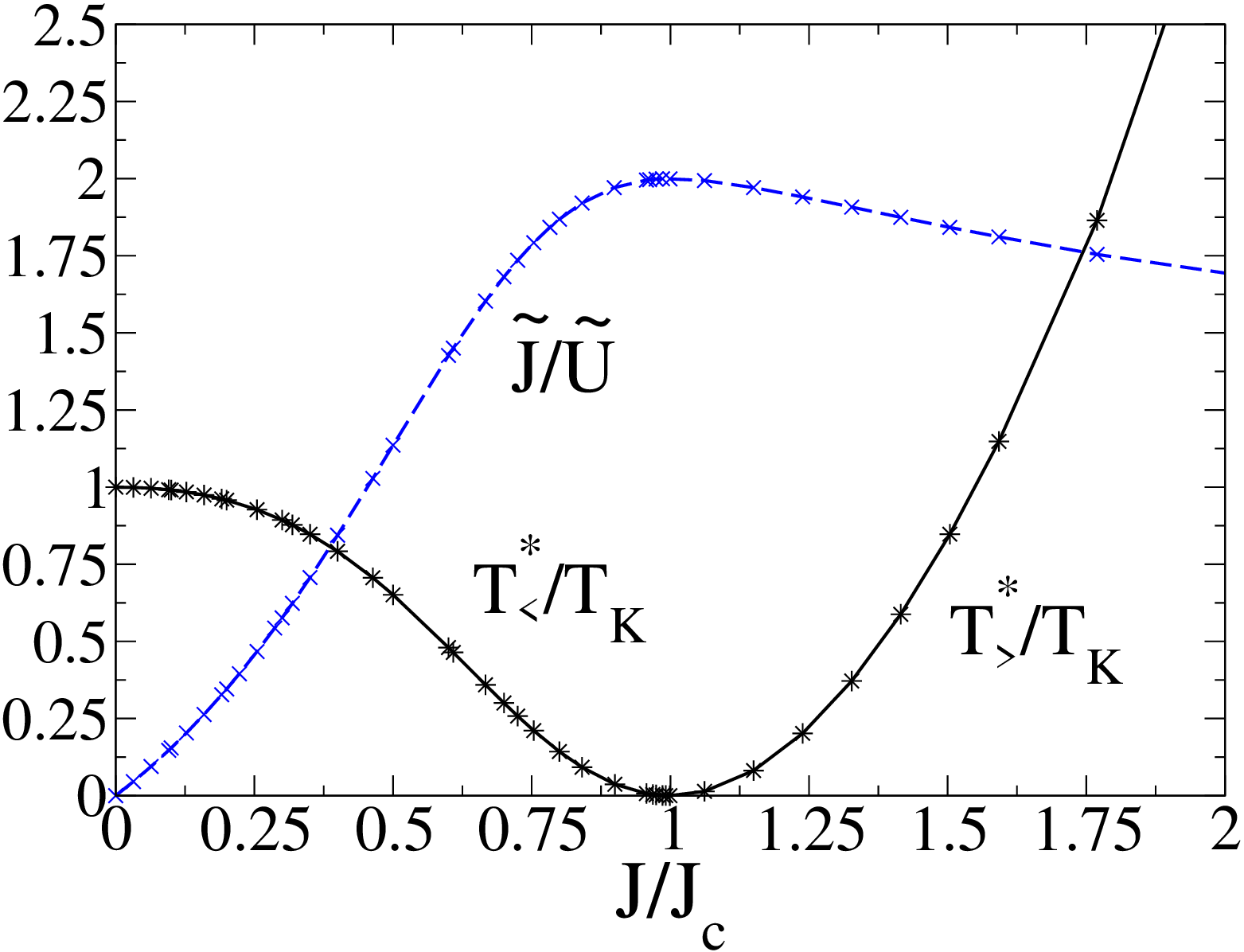}
     \caption{(Color online) A plot of the renormalized parameters,
 $T_<^*/T_{\rm K}$, $T_>^*/T_{\rm K}$ 
 (stars) $\tilde J/\tilde U$ (crosses), as a function of $J/J_c$  for  $U/\pi\Delta=5$.
} 
     \label{rp_ratios}
   \end{center}
 \end{figure}
 \noindent

\noindent
\begin{figure}[!htbp]
   \begin{center}
 \includegraphics[width=0.2\textwidth]{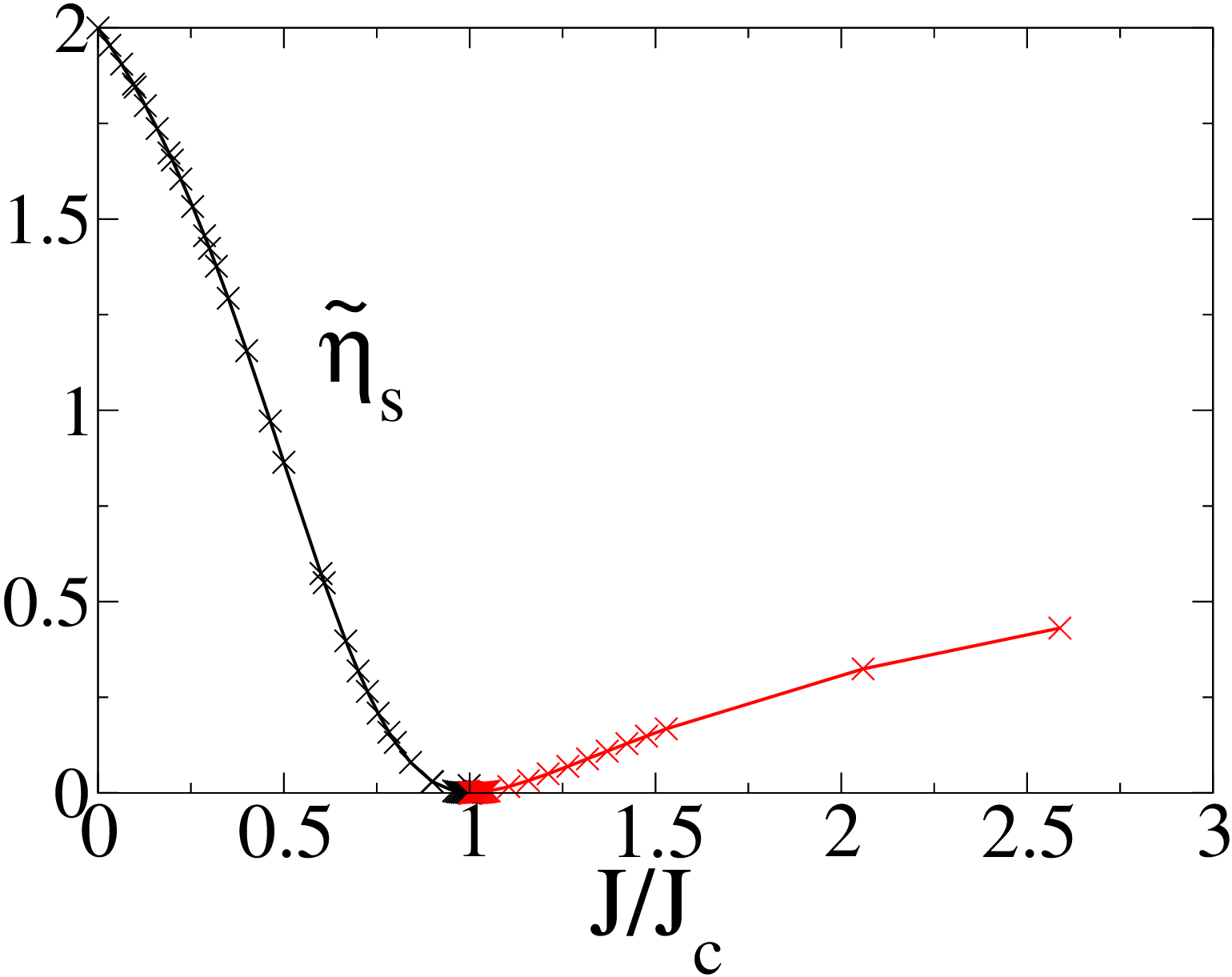}
     \includegraphics[width=0.21\textwidth]{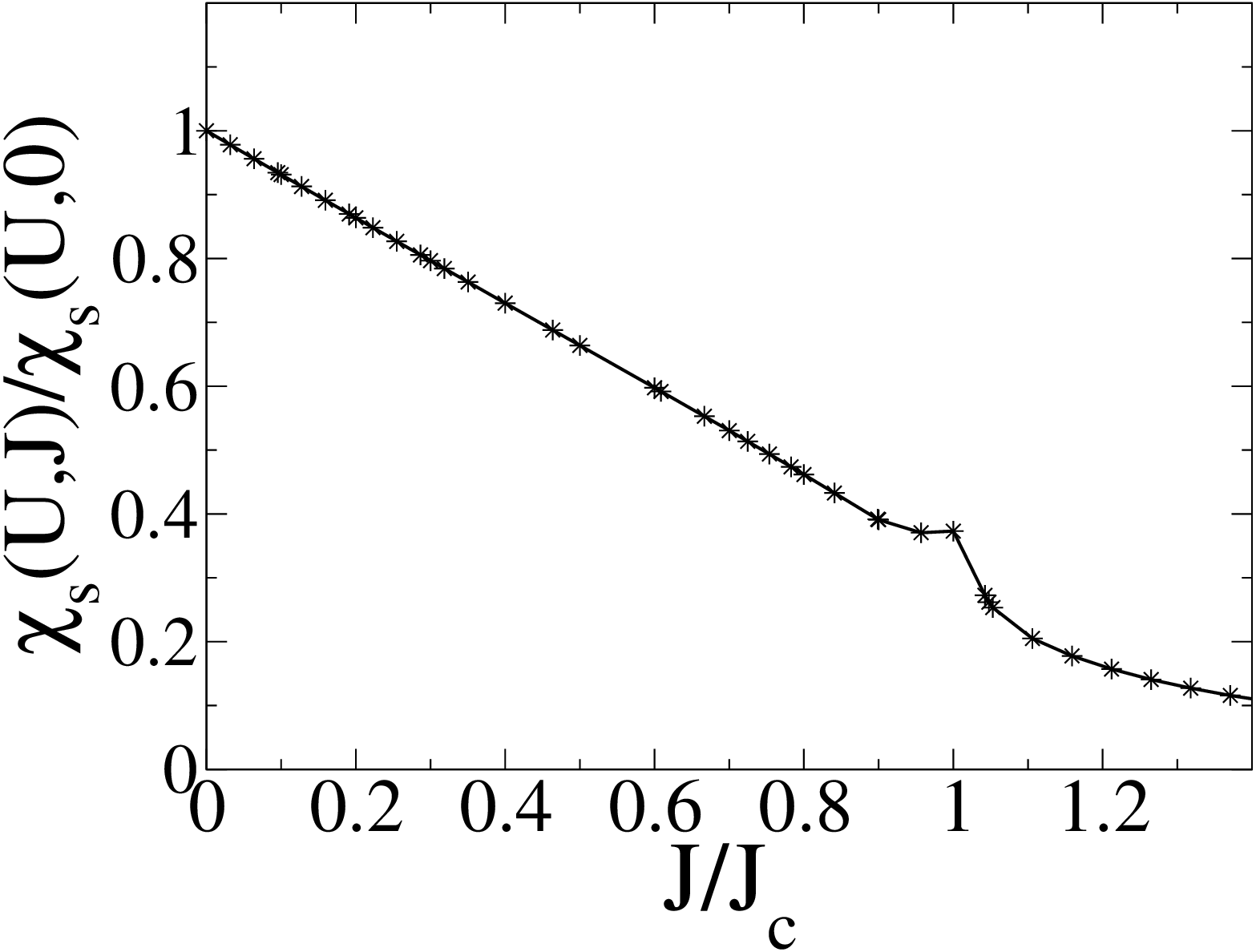}
     \caption{(Color online) A plot of the Wilson ratio $\tilde\eta_s$
(left) and the spin susceptibility ratio,
$\chi_s(U,J)/\chi_s(U,0)$, as a function of $J/J_c$ for $U/\pi\Delta=5$.}
     \label{chisg}
   \end{center}
 \end{figure}
 \noindent
 At $J=J_c$ there is a discontinuous change in the NRG fixed point
as the two impurities decouple from the conduction band on the lowest
energy scale. The phase shift $\delta$ changes from  $\pi/2$ to $0$
due to the singularity developing in the self-energy of the impurity Green's
functions. For $J>J_c$, $z=0$ and the previous analysis based on the assumption of
 analyticity of the self-energy at $\omega=0$ breaks down. However,
in this regime the low energy behavior still corresponds to a Fermi liquid.
We can retain Eqns (\ref{chi}) and  (\ref{gam}) as a description of a local
Fermi liquid 
and  treat the  first conduction site in the NRG chain as an effective
impurity. We can then derive effective renormalized parameters as for $J<J_c$,
but we have to take into account that the hybridization is now to a modified conduction chain.
The renormalized quantity $\tilde \Delta$ is nolonger equal to $z\Delta$,
so in using it to define an energy scale $T^*$ we distinguish it from the
values for $J<J_c$, by $4T^*_>=\pi\tilde \Delta$.\par
In Fig. \ref{rp_ratios} we show the results for the renormalized parameters
over the range through the transition point as a function of $J/J_c$
for $U/\pi\Delta=5$. It can be seen that $\tilde J
/\tilde U$ is continuous through the transition and takes the predicted
value 2 at $J=J_c$. The curves for   $T^*_<$ and  $T^*_>$ approach the 
critical point in a similar way proportional to $(J-J_c)^2$ so that it seems reasonable to identify them as a single energy
scale $T^*$. The results shown  were found to be universal in the strong
correlation regime $U/\pi\Delta>3$.\par 
The value of $\tilde\eta_s$, which is the Wilson ratio, is shown as a function of
$J/J_c$  in Fig. \ref{chisg} (left). The corresponding spin susceptibility
$\chi_s=4\mu_{\rm B}^2\tilde\eta_s/\pi\tilde\Delta$, is difficult to determine
precisely in the immediate region of the critical point as both $\tilde\Delta$
and $\tilde\eta_s$ tend to zero in this limit. The value  
of $\tilde\eta_s$ depends on the difference between the renormalized
parameters which are very small in this regime, so any errors in the determination of the
parameters become significant. As $J$ is increased to
$J/J_c=0.95$ there is an almost linear decrease with $J/J_c$. For $J/J_c>1.05$
the susceptibility appears to fall off more slowly with increase of $J/J_c$. An interpolation between
the two regimes, as  shown in Fig.   \ref{chisg} (right), suggests that
there could be a peak at the critical point $J=J_c$, but it is very sensitive to
the range of the interpolation regime.   
\par

\begin{figure}[!htbp]
   \begin{center}
     \includegraphics[width=0.233\textwidth]{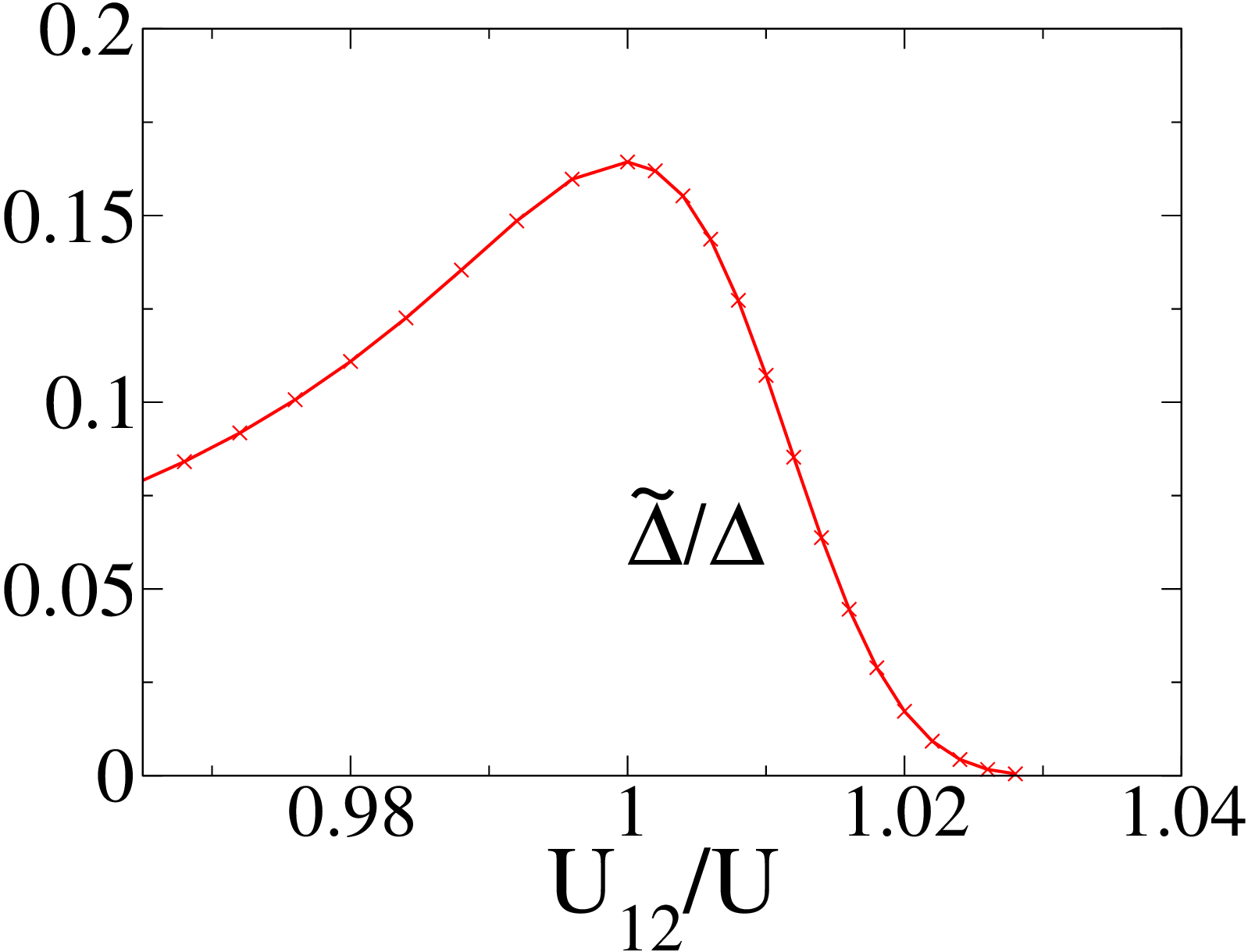}
  \includegraphics[width=0.215\textwidth]{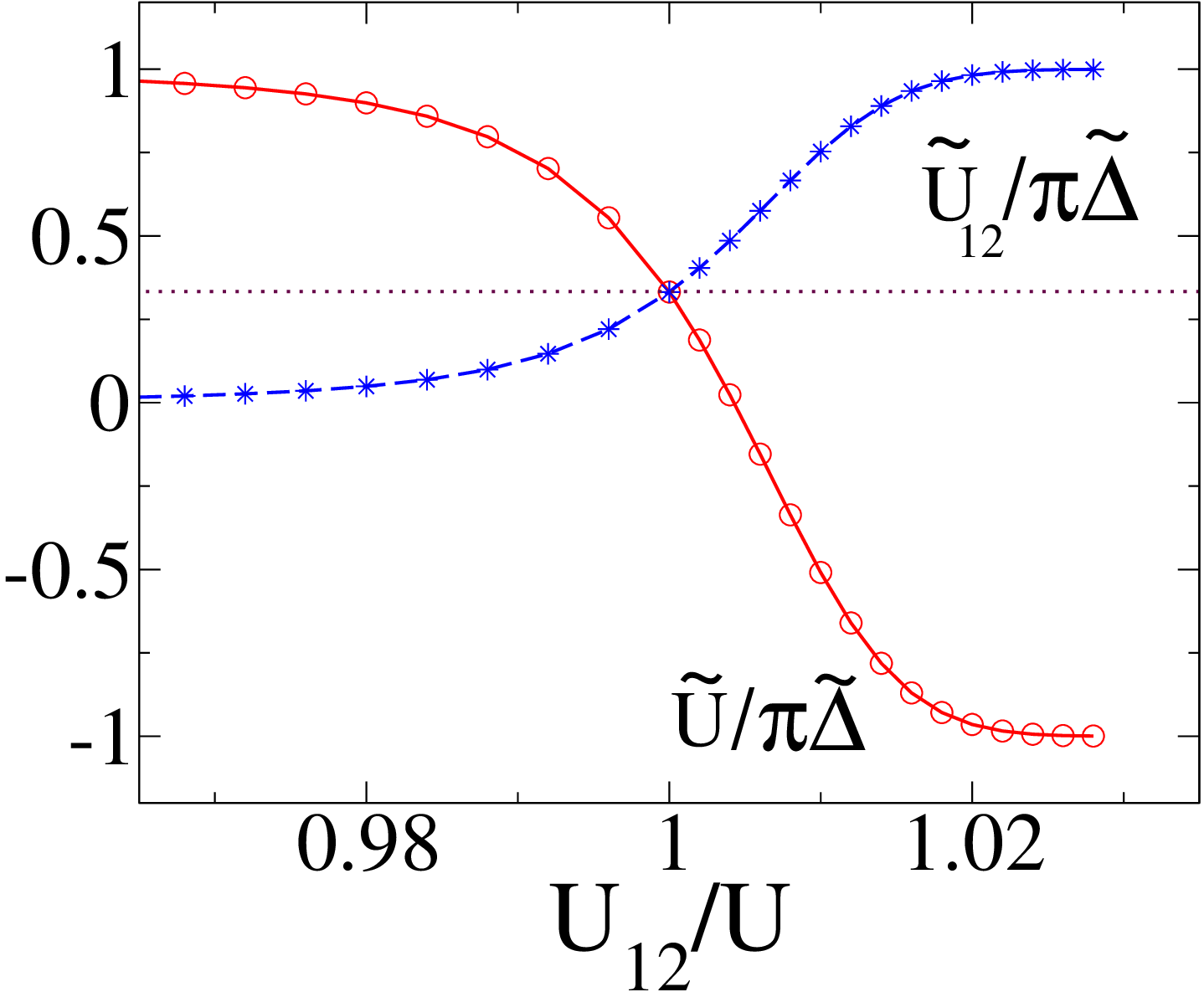}
     \caption{(Color online) A plot of $ \tilde\Delta/\Delta$ (left)
and $\tilde U/ \pi\tilde\Delta$ (circles)
and $\tilde U_{12}/ \pi\tilde\Delta$ (stars) (right)
as a function of $U_{12}/U$ in the approach to the charge order transition for  $U/\pi\Delta=5$.}
     \label{rdelU12}
   \end{center}
 \end{figure}
 \noindent

 Results for the transition to the local charge ordered state for the 
model with $U_{12}\ne 0$ and $J=0$ are shown in Fig. \ref{rdelU12}
for $U/\pi\Delta=5$
as a function of $U_{12}/U$. As $U_{12}$ is increased there is a point   
 $U_{12}=U$ with SU(4) symmetry where  $\tilde U_{12}=\tilde
 U=\pi\tilde\Delta/3$, which is predicted from the fact that for
large $U$ both the spin
and channel fluctuations are suppressed. The critical point occurs
for $U>U_{12}$, where $\tilde\Delta\to 0$ implying $z\to 0$ and the disappearance
of the resonance at the Fermi level. 
 There is a rapid reduction in  $\tilde
U/\pi\tilde\Delta$ from the SU(4) point to a value -1 at the transition and a commensurate
increase in the value of  $\tilde
U_{12}/\pi\tilde\Delta$ to the value 1, in complete agreement with the
predictions based on Eqn. (\ref{chi}).  As the quantum critical point (QCP) is approached we
again have a single energy scale $T^*$ such that $\tilde U_{12}=-\tilde
U=4T^*/\pi$, $\tilde J=0$, which are also found to apply  for
the model with finite $J$ as $\tilde J/\pi\tilde\Delta\to 0$  at the transition.\par

In summary, we see that universality appears on the approach to the quantum critical points, such that
 the renormalized parameters specifying the low energy behavior can be
 expressed in terms of  a single energy scale $T^*$. At the critical points $T^*\to 0$   the quasiparticle weight factor
$z\to 0$ and the spectral density of the impurity levels at the Fermi level goes to zero. The quasiparticle
 interactions are equal  or greater than the renormalized effective band width
 $\pi\tilde\Delta$, as in the  strong correlation regime. The arguments used here should be generally applicable
to models of heavy fermions as all the susceptibilities in Fermi liquid theory at $T=0$ take the form $\chi_\alpha\propto\tilde\rho(0)\tilde\eta_\alpha $
where $\tilde\rho(0)$ is the  density of states of the non-interacting quasiparticles at the Fermi level and $\tilde\eta_\alpha $ is a factor which depends on the interactions between the quasiparticles. If the specific heat coefficient,
which is proportional to  $\tilde\rho(0)$ diverges at the QCP and the susceptibility $\chi_\alpha$ is finite, then
$\tilde\eta_\alpha=0 $ gives a constraint on the quasiparticle interactions.  The emergence of a single low energy scale $T^*$ means that the low energy dynamic response functions would have the form $F(\omega/T^*,T/T^*)$.
This would  be a natural precursor of $\omega,T$ scaling because as $T^*\to 0$, it would be expected to go over to a form $T^\gamma f(\omega/T,1)$. Calculation of the renormalized parameters from the NRG within a dynamical mean field theory for a lattice model would require
the self-consistent solution of the effective band conduction density of states for a two band model.\par

Some recent interesting experiments have set out to examine the QCP in a two impurity Kondo model by measuring the
current between a cobalt atom on an STM tip and a cobalt atom on a metal surface \cite{BZD11}. The results are given as a function of the
bias voltage so are under non-equilibrium conditions. There is a direct hybridization term between the cobalt atoms
which is not in our model but could be included. Once the renormalized parameters have been determined it is possible to calculate
precisely the differential conductance at low bias voltage,  using the Keldysh version of the renormalized perturbation theory. This approach could be used to calculate the onset of the splitting of the Kondo resonance seen in these experiments
in a similar way to the calculation of the onset of the splitting in a magnetic field in a quantum dot \cite{HBO05}.

{We thank Akira Oguri and Johannes Bauer for helpful discussions. Two of us
  (DJGC and ACH) thank the EPRSC for support (Grant No. EP/G032181/1).
%% The numerical calculations were partly carried out on SX8 at  YITP in Kyoto University.
\bibliography{artikel}

\end{document}